\documentstyle[amsfonts,aps]{revtex}    
\begin{document}
\twocolumn 
\draft

\title{More on``Atomic motions in the crystalline 
Al$_{50}$Cu$_{35}$Ni$_{15}$ alloy''}

\author{Gerrit Coddens} 
\address{Laboratoire des Solides Irradi\'es,\\
C.E.A./C.N.R.S./Ecole Polytechnique, F-91128-Palaiseau CEDEX, France}

\date{\today}

\maketitle
\widetext

\begin{abstract}

We refute recent claims that 
ultrafast atomic jumps
as observed in quasicrystals (QC) could be called phasons in 
many crystalline alloys by pointing out that
there is a genuine conceptual difference between the hopping dynamics
in an imperfect crystal containing a substantial number of vacancies, and
the hopping dynamics due to phason motion in QC.

\end{abstract}

\narrowtext

Recently we reacted\cite{comment} to comments by Dahlborg et 
al.\cite{Dahlborg,Stuttgart}
on work of phason hopping in quasicrystals (QC). 
We want to remove a number of misunderstandings that could result from 
the reply\cite{reply} of these authors and Dolin\v{s}ek. 

It would appear as though we made a severe mistake 
by calling the sample 
a B2-phase rather than a $\tau$-phase.
In one of the references we cited 
\cite{Stuttgart} the authors stated: 
 ``The structure of this alloy is of B2, CsCl type with 
lattice parameter $a=$ 32.9 {\AA}.''
Admittedly, the term B2-phase is outdated 
and should be replaced 
by CsCl-type
phase. It has usually a broad  stability 
range and may contain also
structural vacancies. If the vacancies order leading to
superstructures then these phases are called 
$\tau$-phases. If the
B2-phase is fully disordered it is usually 
called $\beta$-phase. 
But it should be clear that an emphasis on such formal aspects of the
presentation of our argument cannot address its contents.
The issue is not how we name the sample, nor 
what its precise structure is,
but that it is biased in having a significant 
structural-vacancy concentration.

Besides a B2 and a $\tau$ phase the authors  call their 
sample also a {\em one}-dimensional QC. If true, 
this would completely undermine 
the pertinence of their sample
as a counter example of an alloy that displays 
ultrafast hopping but is not a QC.
Let us note in this respect also a tacit change of issues
between the original paper (ultrafast atomic jumps are a trivial
feature of all Al-based ternary alloys)
and the reply  (atomic jumps can be called
phasons in many other solids than QC).
In other words, the authors now claim that the concept of a phason
(and of a QC) could be generalized to such an extent
that it becomes a trivial object.
They base these claims on a geometrical construction
by Steurer.\cite{Steurer}

Steurer's paper aims at introducing a 1-1 mapping between a QC and a
crystal. Before analysing the argument we want to point out
that this starting point is not completely rigorous since 
mathematically spoken it is impossible
to have such a 1-1-mapping, as has been
conclusively shown by Duneau and Oguey.\cite{remark}  
To express that we concede to neglect this problem we could state\\ 

\vspace{2.5cm}

\noindent 
(admittedly quite vaguely) that we allow for ``imperfect''
crystalline structures, although in principle
this should require detailed further specification. 
Steurer's construction consists then in defining 
two mappings: one, which is 1-1,
between a  QC and
an average structure (AS) which is a (highly imperfect) 
periodic lattice with 
a well-determined
structural-vacancy concentration, and one between the 
CsCl-type structure
and the AS structure. In the case of an icosahedral QC 
(the Amman tiling), the second mapping can be the identity
mapping. While it is not wrong that this way a QC can 
be put into registry
with a {\em highly imperfect} CsCl-type structure with a 
uniquely determined 
structural-vacancy concentration 
(e.g. $(3\tau - 4)/(3\tau + 1)$ for 
the construction starting from a decagonal QC),
it must be clear that on the contrary the reversal of this 
is just not true: 
An (imperfect) periodic structure which is the AS 
of a incommensurately 
modulated structure (IMS) is not itself an IMS, as the 
authors suggest.
(Every perfect periodic structure can be used as the 
basic or average lattice
for the definition of some IMS, which of course does not
imply that all periodic structures would be IMS).
Moreover, none of the cubic structures
that have a different structural-vacancy 
concentration than the magic one, can 
be ``considered'' as a QC according to the philosophy of the authors. 
If the vacancy concentration in the AlCuNi sample
were to be the magic one, then it would even be more biased. 

Let us now discuss the generalized ``phasons'' of the authors.
The second mapping does not use hyperspace, such that the nature of the
phasons invoked by the authors must be obtained by transposing
 the picture of
the phasons under the first mapping, using
the second mapping as a dictionary. 
But the first mapping represents the QC as an 
IMS. 
A shift of the superspace cut
in an IMS produces small atomic displacements of the atoms around 
the average positions,
but {\em{leaves the positions of the AS unchanged}} (see e.g. Fig. 1b of
 reference [5a]). 
The images of these average positions, which are 
the atoms of the cubic phase,
 will thus not move eighter.
Hence vacancy diffusion on the CsCl-type
structure does not correspond to phasons (as {\em{the phasons 
remain restricted to the QC and do not 
lead to any atomic
motion on the cubic lattice at all}}), but to true vacancy diffusion 
on the AS (and the QC).
Intersticialcies on the CsCl-type structure are not addressed 
in Steurer's construction.

As the reader can inspect on the Figure just mentioned, 
for large enough amplitudes
of the translation of the cut in hyperspace 
the phason modes of the QC in the IMS setting correspond 
to a gradual (continuous)
shift of the position up to a maximal distance
followed by a (discontinuous) jump back to the starting position.
That is a meaningless concept in the lattice dynamics of a 
cubic structure
(and a QC).

There are no or very few data on hopping rates in ternary alloys.
The claims of the authors that ultrafast hopping would be common
in ternary Al-based alloys can therefore
not be refuted by showing ample experimental 
evidence from the literature
that would prove it wrong.
However, there is more to scientific credibility of a statement
than the absence of experimental information contradicting it.
The statement of the authors is
based on the extrapolation of a single data point collected
on a heavily biased sample. Pointing this out
 should be sufficient to dismiss the entire argument.
There is no further charge of proof on our shoulders 
({\em Actori incumbit probatio}).
Still, for further appraisal of the 
unusual, {\em ad hoc} character
of this extrapolation,
we showed that the jump rates observed in QC are almost a factor of 
100 faster than in all other metallic alloys studied, by 
citing the examples with 
the fastest hopping  rates ever observed in {\em any} non-QC alloy.
(That these examples just happened to be B2 phases is here not part of
the argument).
By replying to this that one cannot compare a ternary phase 
to a binary phase, the authors mispresent the issues.
 
They also mispresent the issues when they claim that we 
would be speculating about the temperature dependence.
If they want to prove that the fast jumps observed in QC are a
common feature of all Al-based ternary alloys, 
then their temperature dependence {\em has
 to be} of the  {\em assisted} type
like in QC. Else the situations in the two samples just do not compare.
Exactly this crucial information about the temperature dependence in
 their own sample is missing, a deficiency we flagged. 
Elsewhere, the authors tacitly assume a {\em normal} temperature
dependence in their sample, by invoking a thermal-vacancy 
mechanism for the fast jumps.

It is, perhaps, not much of a ``surprise''
that we take great discomfort with a
21-pages Comment on our work having been presented 
as a regular article,
and the inordinate delay of what just should have been
our right of reply and the final word in the discussion. 

Finally, we would like to signal an erratum
that sneeked into our paper due to editorial changes during the 
production process.
We wanted to quote Hf and $\beta$-Ti ``show hopping rates two 
orders of magnitude slower''
rather than ``show self-diffusion two orders of magnitude slower''.

We would like to thank Prof. W. Steurer for providing  us with 
some accurate definitions, and a copy of a paper [5a]
we were unable to get access to.

\end{document}